SOLIDS
AND LIQUIDS

# Fano Interference at the Excitation of Coherent Phonons: Relation between the Asymmetry Parameter and the Initial Phase of Coherent Oscillations

O. V. Misochko* and M. V. Lebedeva

*Institute of Solid State Physics, Russian Academy of Sciences, Chernogolovka, Moscow oblast, 142432 Russia*
*e-mail: misochko@issp.ac.ru
Received September 10, 2014

**Abstract**—The theoretical assertion that the Fano asymmetry parameter and the asymptotic initial phase of a harmonic oscillator interacting with a continuum are interrelated is experimentally verified. By an example of coherent fully symmetric $A_{1g}$ phonons in bismuth that are excited by ultrashort laser pulses at liquid helium temperature, it is demonstrated that, for negative values of the asymmetry parameter, the asymptotic phase increases as the modulus of the parameter decreases.

## 1. INTRODUCTION

Resonances are widespread in nature and are encountered in various physical problems, starting from mechanical and electromagnetic oscillators in classical physics to various excitations in quantum systems. In solid state physics, the resonances are associated with quasiparticles, whose typical representatives are phonons, band electrons, and plasmons. Most often, a resonance or its cross section is expressed by a Lorentzian function or a symmetric Breit–Wigner profile [1],

$$\sigma(\varepsilon) \propto \frac{1}{1+\varepsilon^2}, \qquad (1.1)$$

where the dimensionless energy $\varepsilon = (E - E_r)/(\gamma/2)$ is measured in units of the half-width of the profile that determines the lifetime and $E_r$ is the resonance frequency. The symmetry of the profile follows from the exponential decay of the state in which the decay probability is independent of time. Resonance may arise as a result of interference between alternative paths connecting the initial and final states of the system in the same way as they arise in the canonical example of a two-slit interference experiment. Just as any interference phenomenon, the cross section in this case includes the contributions from both constructive and destructive interference. These interference phenomena manifest themselves in the asymmetric profile of the scattering cross section, which was first theoretically considered in this context by Fano [2, 3], who solved the problem of configuration interaction formulated by Majorana [4] in the case of a discrete level and a continuum:

$$\sigma(\varepsilon) \propto \frac{(q+\varepsilon)^2}{1+\varepsilon^2}. \qquad (1.2)$$

The Fano profile illustrates that, due to interaction $V$, the continuum acquires a structure in the range of energies determined by the inverse lifetime of a modified discrete level. In contrast to Eq. (1.1), in which the cross section depends on the energy and lifetime of the resonance, expression (1.2) includes a third parameter $q$, called the asymmetry parameter. The dynamics of any system depends on this parameter much stronger than on the resonance energy and/or lifetime, which, in the presence of a continuum, are renormalized by the interaction $V$:

$$E_d = E_r - V^2 R(\varepsilon), \quad \Gamma = \gamma + V^2 \rho(\varepsilon). \qquad (1.3)$$

Here $\rho(\varepsilon)$ is the density of states of the continuum and $\pi^{-1} R(\varepsilon)$ is its Hilbert transform. The asymmetry parameter $q$ is defined by the ratio of two amplitudes related to interference: the amplitude $A_d$ of transition to the discrete state and the amplitude $A_c$ of transition to the continuum, as well as by the coupling constant $V$ between the discrete level and the continuum, which is responsible for the renormalization of the discrete state,

$$q = \frac{V(A_d/A_c) + V^2 R(\varepsilon)}{\pi V^2 \rho(\varepsilon)}. \qquad (1.4)$$

Since constructive interference leads to an increase in the cross section on one of the slopes of the profile, while destructive interference manifests itself as a decrease on the second slope, the resonance takes an asymmetric shape. The energy scheme of the Fano interference is demonstrated in Fig. 1a. If there is no excitation of the continuum ($A_c = 0$, $|q| \rightarrow \infty$), then the spectral line has a symmetric Lorentzian shape with the half-width due to the lifetime $\Gamma^{-1}$ of the discrete level modified by the interaction. If there is no excitation of the discrete level ($A_d = 0$, $q = 0$), then there is a symmetric minimum in the spectrum, which is called an antiresonance. When both amplitudes are different from zero, the constructive and destructive interference for energies greater or less than the energy of the modified discrete state, respectively, result in an asymmetric line with a maximum at $\varepsilon = 1/q$ and a minimum at $\varepsilon = -q$, as is shown in Fig. 1b. Here the pole and the zero of the scattering amplitude in the complex plane correspond to the resonance and antiresonance, respectively. The antiresonance phenomenon is the most unique property of the Fano interference, which means that, at some frequency, the state of the system cannot be changed.

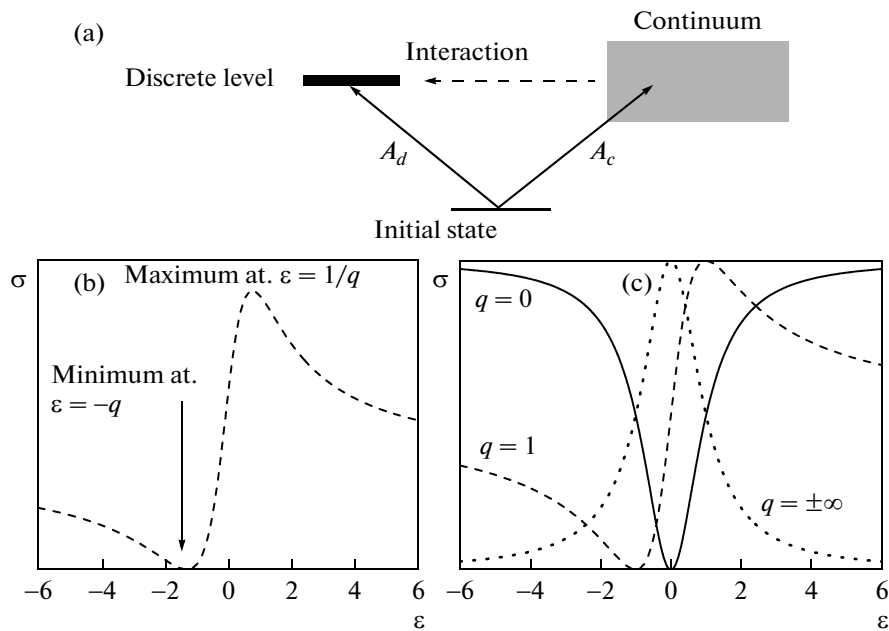

**Fig. 1.** (a) Energy diagram schematically illustrating the Fano interference. (b) Normalized cross section as a function of dimensionless energy for several values of the asymmetry parameter $q$. The curves for positive values are obtained by inverting the abscissa axis. (c) Positions of extrema determined by the zero and the pole of the Fano profile.

In fact, the Fano interference is one of paradigms of modern physics that allows one to investigate the fine details of interaction and the internal structure of quantum objects. It is well known that Fano's paper [3] is one of the most frequently cited papers in modern physics, which points to a rather general character of the model that is implemented in a wide class of phenomena. The model is encountered in various problems of nuclear, atomic, and molecular physics, as well as in many problems of solid state physics and optics [5]. The number of publications devoted to the Fano interference is enormous, and it is impossible to list all of them. We just note that, since, in the experiments on the study of this phenomenon that are carried out in the frequency domain, one used, as a rule, square-law detectors, which are insensitive to the phase of the radiations used, the phase relations responsible for the interference behavior have remained poorly studied so far. The situation has drastically changed with the appearance of ultrashort laser pulses whose duration reaches the attosecond range.[1] The use of such pulses allows one to time-resolve the dynamics of many processes and, in particular, in the cases of phonons, to get access not only to the displacement amplitudes of atoms but also to their phases [6, 7]. For example, in the case of excitation of lattice modes in a crystal by femtosecond laser pulses, in the last decades, the dynamics of the Fano resonance of an optical phonon coinciding in energy with the electron continuum in semiconductors [8–10], semimetals [11, 12], metals [13, 14], and high-temperature superconductors [15] has been investigated. The dynamical spectroscopy method, which allows one to create coherent states of lattice modes in the case of femtosecond laser pulses, provides, compared with conventional spectroscopy, additional information that may be useful to get a deeper insight into the Fano interference.

## 2. FANO INTERFERENCE AND COHERENT PHONONS: A THEORETICAL ASPECT

In the experiments on the study of coherent phonons by the ultrashort laser pulse pump–probe method, the spectrum of excited oscillations exhibits, under certain conditions, characteristic features in the form of a Fano profile. The asymmetry parameter defined in the spectral range depends on temperature, doping level, the phonon energy, and the intensity of laser pulses [7–15]. The emergence of such features suggests that, just as in the original Fano problem [3], they arise due to the weak interaction between a discrete phonon mode and the continuum of lattice and/or electron states of the crystal. If this conjecture turns out to be correct, one can learn rather a lot about the character of interaction between coherent phonons and other degrees of freedom of the crystal.

The application of the Fano model in the physics of coherent phonons is also justified from the viewpoint of its possible relation to the mechanism of excitation of coherent lattice modes. It is well known that, in transparent crystals, coherent vibrations of the crystal lattice start after the impulsive excitation from the zero displacement of atoms, i.e., they have the form of a decaying sinusoidal function, whereas, in opaque crystals, the displacement of atoms from a shifted equilibrium position is maximal at the initial instant of time, and the vibrations have the form of a decaying cosine function. It is commonly assumed that this is associated with different mechanisms of excitation of vibrations [6, 7]. In transparent crystals, the matter concerns the impulsive stimulated Raman scattering [16], whereas, in opaque crystals, one deals with the displacive mechanism [17]. Attempts to unify these two limits within a single Raman model were made long ago [18]; however, a number of experimental facts [19] such as the initial phases and the ratio of amplitudes of fully symmetric and low-symmetry lattice modes did not perfectly fit this unified model. In [20], the author made another attempt to link the two above mechanisms and explain the initial phase of excited vibrations within the Fano model.

The author of [20] suggests that oscillations of coherent phonons are weakly coupled to the continuum of electron oscillators of the crystal. The initial phase of the excited oscillations is determined by where the energy of the exciting pulse is deposited during the excitation. According to this author, if a pulse excites the continuum, then the oscillations of an optical phonon mode follow the cosinusoidal law, whereas, if the whole excitation energy is deposited directly to the optical phonon, then the oscillations start as a sinusoid. Naturally, intermediate cases are possible when both the continuum and the discrete oscillator are excited and the initial phase of oscillations turns out to be intermediate between sine and cosine. Note that the initial phase mentioned in [20] is understood as the phase of harmonic oscillations of the discrete oscillator, set up in the system at large times, which is extrapolated to zero time. In this case, the zero instant of time corresponds to the time when impulsive forces act on the system. In order to avoid confusion, we will call the initial phase thus defined the asymptotic initial phase.

The original model of Fano [3] was formulated as a quantum-mechanical problem. The author of [20] considered a classical analog of this quantum-mechanical problem, complementing the frequency analysis of the problem by the analysis of the dynamics in the time domain. Therefore, the comparison of the results of [20] with the corresponding results of Fano [3] are of independent interest.

Consider, first of all, the most general, purely mathematical, aspect of the problem. Suppose we have an oscillation spectrum in the form of a Fano profile. In [3], this profile describes the probability of excitation of the eigenstates of a quantum-mechanical system under the transition operator from a certain given initial state. The probability is a quadratic function of amplitude; therefore, to determine the spectrum of the transition probability amplitude, one should take the square root of this quantity. In the classical problem [20], the situation is completely analogous: a Fano profile describes the spectrum of energy dissipation, $S(\omega)$, in a system subjected to external harmonic forces of given circular frequency $\omega$. In this case, the whole set of coupled oscillators performs forced harmonic oscillations with the same frequency but different amplitudes and phases. In the stationary case, energy dissipation due to the damping of individual oscillators is compensated for by the work of external harmonic forces acting on the system.

Let us pass from the consideration of individual oscillators to the consideration of the set of normal oscillations of the coupled system of oscillators, i.e., from the set of interacting oscillators to the set of independent modes each of which has its own resonance frequency, amplitude, phase, and damping $\gamma$. These modes represent an analog of eigenstates of a quantum-mechanical system, and their oscillations under external forces are analogous to transitions in a quantum-mechanical system under an external perturbation. Since the main contribution to the response to an external action is made by the modes whose eigenfrequencies are close to the frequency of the external driving forces, the dissipation power averaged over a period is proportional to the squared amplitude $x_0$ of oscillations of the oscillator,

$$S(\omega) = \frac{1}{T}\int_0^T \gamma \dot{x}\dot{x}dt = \gamma\langle \dot{x}^2\rangle_T \propto x_0^2,$$

and, to find the oscillation amplitude, one should also extract a square root. Thus, suppose that, as a result, we have, near a certain resonance frequency $\omega_0$, a dissipation power spectrum $S(\omega)$ in the form of a Fano profile, i.e.,

$$S(\omega) = |A|^2 \frac{(q+\varepsilon)^2}{1+\varepsilon^2}, \qquad (2.1)$$

where $\varepsilon = (\omega - \omega_0)/(\Gamma/2)$ is dimensionless energy and $A$ is a normalization constant. In this case, the spectrum of oscillation amplitudes of normal modes $F(\varepsilon)$ has the form

$$F(\varepsilon) = A\frac{q+\varepsilon}{\varepsilon+i} = A\frac{q+\varepsilon}{\sqrt{1+\varepsilon^2}}e^{i\Delta}, \quad \cotan\Delta = -\varepsilon, \quad (2.2)$$

where the phase $\Delta$ of the Fourier coefficient coincides with the relative phase of the eigenstate of the Hamiltonian after the interaction between the discrete state and the continuum of states in the Fano problem is taken into account [3].

The time response of the system is obtained from (2.1) by the Fourier transformation,

$$f(t) = A\exp(-i\omega_0 t)\exp(-\Gamma t/2)$$
$$\times \left[i\delta(t) + \frac{\Gamma}{2}(q-i)\theta(t)\right] \quad (2.3)$$
$$= A\exp(-i\omega_0 t)\exp(-\Gamma t/2)\left\{i\delta(t) + \frac{\Gamma}{2}\sqrt{q^2+1}e^{i\varphi}\theta(t)\right\},$$

where $\tan\varphi = -1/q$, $\theta(t)$ is the Heaviside function, and $\delta(t)$ is the delta function. In this case, the relation between the phase and the asymmetry parameter is given by the expression

$$\varphi = \arctan\left(-\frac{1}{q}\right), \quad (2.4)$$

and the time response represents damped oscillations with resonance frequency $\omega_0$, for which the phase shift $\varphi$ at the resonance frequency $\omega_0$ is due to the presence of the contribution proportional to $\varepsilon$ in the spectrum (2.2) of the response, which is attributed to the excitation of the continuum of oscillators coupled to the discrete oscillator. In the limit as $|q| \to \infty$, this shift tends to zero. Notice that the phase shift $\varphi$ is different from the phase shift $\Delta$ of the Fourier coefficients.

Thus, we can formally replace the original system of coupled oscillators by a single damped oscillator whose initial phase is completely determined by the parameter $q$. The eigenfrequency of this oscillator is slightly different from the original eigenfrequency of the discrete oscillator due to the effect of the continuum. Such a substitution is justified by the closeness of the properties of this damped oscillator to the observed properties of coherent phonons excited by a short laser pulse.

Now, let us analyze the results of [20] in more detail. As pointed out above, the author of [20] passed from the original quantum-mechanical Fano problem to the classical equations of motion of oscillators. He showed that the spectral form of the Fano profile is preserved and the quantities that define this profile are analogous to the corresponding quantities of the quantum-mechanical calculation. The oscillators that form a continuum in [20] possess damping. The introduction of this damping turns out to be essentially necessary because the imaginary part of the work of external forces, which is proportional to energy dissipation, vanishes in the absence of damping. Recall that, in the Fano problem, the original analysis is carried out for the eigenstates of the Hamiltonian that do not decay in time, and the damping of a state of the discrete oscillator that is produced at the initial instant of time and is not an eigenstate of the Hamiltonian occurs due to the dephasing of the eigenstates of the problem that are the components of this state. After obtaining the frequency dependence of energy dissipation in the form of a Fano profile, the author of [20] passes to the limit $\gamma_k = \gamma \to 0$, in which the expressions for the parameters of the profile almost coincide with their quantum analogs.

When considering the time response of the system to short pulsed forces, the author of [20] used the approximation of weak coupling between the discrete oscillator and the continuum. As a result, he obtained the following asymptotic expression for the amplitude $Q$ of oscillations of the discrete oscillator as $t \to \infty$:

$$Q(t) = \left(\hat{F}_Q(\omega_0) + \mathrm{P}\sum_k \frac{c_k}{2\omega_k}\frac{\hat{F}_k(\omega_0)}{\omega_0-\omega_k}\right) \quad (2.5)$$
$$\times \sqrt{1+\frac{1}{q^2}}\cos(\omega_0 t - \arctan q),$$

where the indices $Q$ and $k$ correspond to the discrete oscillator and the continuum, respectively, and the symbol P denotes the Cauchy principal value integration. Thus, the asymptotic phase of the discrete oscillator turns out to be uniquely related to the Fano parameter:

$$\varphi = -\arctan q. \quad (2.6)$$

Thus, since the arctangent is an odd monotonically increasing function, the asymptotic phase increases as the asymmetry parameter decreases [20]. For zero phase, i.e., when oscillations are cosinusoidal, the whole energy is transferred to the states of the continuum, and the spectrum should exhibit a Fano antiresonance.

If we compare the result of [20] (formula (2.5)) with Eq. (2.3), it is evident that, in the first case, the oscillations of the discrete oscillator do not decay in time, whereas, according to (2.3), they should decay. Moreover, the functions (2.4) and (2.6) for the phase are radically different. Therefore, one can raise the following objections against the approach to the problem used in [20]. First, the limit $\gamma_k = \gamma \to 0$ should not lead to the vanishing of damping of the discrete oscillator, because the latter occurs in the Fano problem due to the dephasing rather than due to energy dissipation. Second, since the value $|q| \to \infty$ corresponds to the resonant excitation of the oscillator under which the Fano profile turns into a Lorentzian curve, the problem should reduce to the excitation of a single oscillator. In this case, the time response is easily obtained by the Fourier transform, which results in a decaying cosinusoidal function, whereas (2.5) yields a sinusoidal function. It is easily seen that the solution obtained by the author of [20] is proportional to the real part of the Fourier component of the Fano amplitude profile at frequency $\varepsilon = 1/q$ at which the Fano curve has a maximum:

$$\mathrm{Re}\left[F\left(\omega = \omega_0 + \frac{\Gamma}{2q}\right)\exp(-i\omega_0 t)\right]$$
$$= A(\omega_0)q\sqrt{1 + \frac{1}{q^2}}\cos(\omega_0 t - \arctan q). \quad (2.7)$$

Thus, we can conclude that, in spite of the fact that the author of [20] was the first who drew attention to the fundamental relationship between the method of initial excitation of the system (either through the discrete oscillator or through the continuum) and the asymptotic initial phase of coherent phonons, his result does not agree with the Fourier transform of the amplitude Fano profile given by Eq. (2.3).

Concerning the substitution of classical equations for the quantum equations of the Fano problem, we should make the following remark. It is well known that the quantum description of a quantum oscillator in the Heisenberg representation is most close to the classical description because noncommuting quantities do not play here any decisive role. Nevertheless, there remain a number of subtle questions when passing to the classical description. When replacing operators by numerical functions, it is implicitly assumed that all oscillators are in coherent states, which are the eigenstates of annihilation operators. This assumption may not hold under some conditions, and then there may be no classical description for a system of interacting oscillators [21]. However, when the continuum in the Fano problem is formed by electron states and the discrete level is related to lattice excitations, the parameter that determines the possibility of transition to the classical description is given by the ratio of the Debye frequency to the Fermi energy [22]. It is the smallness of this parameter that admits the use of the kinetic equation, similar to the case of the Drude conductivity problem.

## 3. FANO INTERFERENCE AND COHERENT PHONONS IN BISMUTH: EXPERIMENTAL RESULTS

As an example of the Fano interference in the case of coherent phonons, consider the response of a bismuth single crystal photoinduced by an ultrashort laser pulse at liquid helium temperature. Bismuth is crystallized into a rhomohedral lattice $A7$ with two atoms in the unit cell. Its structure can be considered as a result of stretching a simple cubic lattice along one of spatial diagonals, accompanied by a relative shift of two adjacent face-centered sublattices [23, 24]. Of the six phonon modes of bismuth, the optical modes $A_{1g}$ and $E_g$ are Raman active and can be coherently excited by ultrashort laser pulses [25, 26]. In this case, fully symmetric $A_{1g}$ phonons of bismuth are formed by out-of-phase displacements of the atoms of the unit cell along the diagonal with respect to which the Peierls shift occurs. These phonons modulate the internal shift, whereas the doubly degenerate $E_g$ phonons modulate a trigonal shift and arise due to the out-of-phase displacements of atoms perpendicular to the trigonal axis.

Fully symmetric $A_{1g}$ phonons in an undeformed cubic lattice correspond to longitudinal acoustic modes, whereas $E_g$ phonons originate from transverse acoustic modes at the $R$ point of the Brillouin zone. Under strong excitation, both coherent phonon modes demonstrate a chirp (dependence on time) of the frequency, which is translated into the frequency domain as the asymmetry of the profiles of lines in the spectra obtained by the Fourier transformation of a time-resolved reflection signal [25, 26]. An internal shift that doubles the volume of the unit cell and removes the degeneracy between $L$ and $T$ points of the Brillouin zone makes bismuth into an insulator (a direct gap of about 30 meV at the $T$ point), whereas a trigonal shift is responsible for the insulator–semimetal transition [27]. The Fermi surface of rhombohedral bismuth consists of three electron and one hole ellipsoids; the extrema of the conduction band are at the $L$ point, while those of the valence band, at the $T$ point of the Brillouin zone [23, 24].

To study the Fano interference, we used a femtosecond laser system consisting of a generator of femtosecond pulses based on sapphire titanate and a regenerative amplifier (with a wavelength of $\lambda = 800$ nm, a pulse duration of $\tau = 45$ fs, and a pulse repetition rate of 250 kHz). The measurements were carried out in a degenerate pump–probe scheme at liquid helium temperature. Laser pulses were split by a beam-splitting plate with tenfold attenuation of the probe pulse intensity compared with that of the pump pulse. The delay of probe pulses with respect to pump pulses was implemented by a delay line with a step motor. Both pump and probe beams were focused by achromatic lenses to regions with diameters of 0.1 and 0.05 mm on the base plane (0001) of the crystal; i.e., the probe size was much less than the size of the pump region, thus allowing one to minimize the effects of inhomogeneity of excitation [28]. Changing the polarization of the electric vector **E** of the driving laser pulse with respect to crystallographic axes, one can initiate the motion of atoms in different directions [29], and, changing the intensity of the laser pulse, one can vary the displacement of atoms from instantaneous equilibrium positions [30]. The experimental information was obtained as a set of normalized difference signals of reflection from the crystal,

$$\frac{\Delta R}{R_0} \equiv \frac{R(t) - R_0}{R_0},$$

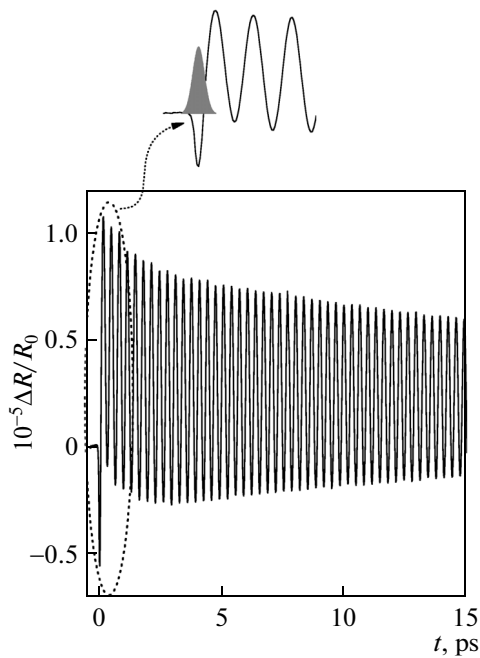

**Fig. 2.** Time-resolved normalized differential reflection $\Delta R/R_0$ of a Bi single crystal at liquid helium temperature in the case of weak pumping ($F = 0.15$ mJ/cm$^2$). The inset shows the initial part of the signal along with the autocorrelation function of a laser pulse.

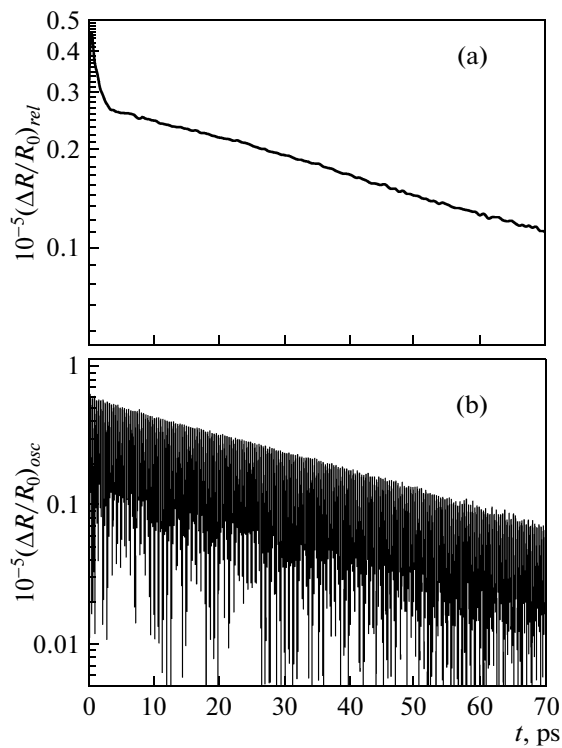

**Fig. 3.** Decomposition of photoinduced reflection into (a) relaxation and (b) coherent components. Each contribution is shown in a semilogarithmic scale, which makes obvious the exponential character of decay of signals.

taken at different instants of time before ($R_0$) and after ($R(t)$) the excitation of the crystal. A detailed description of the experimental conditions is given in the earlier papers [28, 30, 31].

In this paper, we investigate fully symmetric $A_{1g}$ phonons of bismuth excited in a linear regime, which is characterized by the proportionality between the displacement amplitude of atoms and the intensity of excitation [30]. First, consider fully symmetric phonons of bismuth under weak excitation (for energy density of a pump pulse of $F = 0.15$ mJ/cm$^2$) when a coherent (oscillating) contribution slightly dominates in the photoinduced response, part of which is shown in Fig. 2. One can easily verify this by decomposing a signal into a coherent and relaxation (nonoscillating) contributions, which is illustrated by Fig. 3. However, under this decomposition, it remains unclear in which contribution the interaction between the discrete (phonon) level and the continuum remains. Therefore, below we will analyze, as a rule, only the total signal.

The coherent, oscillating contribution of the response is due to the motion of atoms of the crystal and can be associated with the discrete level in the Fano problem, whereas the relaxation contribution is responsible for the state of the continuum whose energy spectrum overlaps with the phonon energy. Under weak excitation, this relaxation contribution decreases biexponentially, the fast component decaying with a characteristic time of about 800 fs, whereas the characteristic lifetime of the slow component being greater than 100 ns (see Fig. 3). After the excitation pulse, the electron system is withdrawn from the equilibrium state and soon after this reaches a state in which the variation of reflection is maximal. Then the system again returns to the equilibrium state, which is reached within a time of slightly less than 100 ps. The oscillating contribution starts from the negative minimum, whose value is less than the oscillation amplitude and the emergence time roughly coincides with the duration of the laser pulse [32, 33]. In [32], this negative minimum was attributed to the effect of an impulsive force of polarization nature, which is proportional to the gradient of the electron density pressure and is separated from the force proportional to the electron density, which leads to an increase in the reflection and provides a basis for the displacement mechanism. The time interval that separates the negative minimum from the first maximum of the signal is not greater than 260 fs, which is much greater than the half-period of fully symmetric phonons and roughly coincides with the half-period of doubly degenerate phonons. For positive time delays, oscillations occur with a period of $325 \pm 5$ fs, which remains constant for small and large delays.

The Fourier transform of the total signal shown in Fig. 4 indicates that the symmetric peak at frequency of 3.06 THz, which coincides with the energy of a fully symmetric $A_{1g}$ phonon, dominates in the amplitude spectrum. The spectral line shape of this peak is symmetric and is well approximated by either a Lorentzian or a Fano profile with large modulus of the negative asymmetry parameter $q$. The initial phase of oscillations is cosinusoidal (is equal to $-\pi$), which can easily be verified either by real-time fitting, the result of which is demonstrated in Fig. 5, or from the phase spectrum shown in Fig. 4. This is also testified by the real part of the Fourier transform of the experimental signal $\Delta R/R_0$, shown in Fig. 6, which is negative and symmetric with respect to the resonance frequency. This part integrates the cosine components of the signal, while its negative part indicates that fully symmetric oscillations start from the negative extremum, which corresponds to the trough of oscillations rather than to their crest or zero. Since the oscillation frequency under weak excitation is independent of time, the asymptotic phase φ coincides with the initial phase of the experimental curve, which is equal to $-\pi$. This coincidence of the asymptotic phase with the experimental one under weak excitation may provide evidence [20] for the fact that the width of the continuum spectrum is large enough, which is also true in the case of silicon [10].

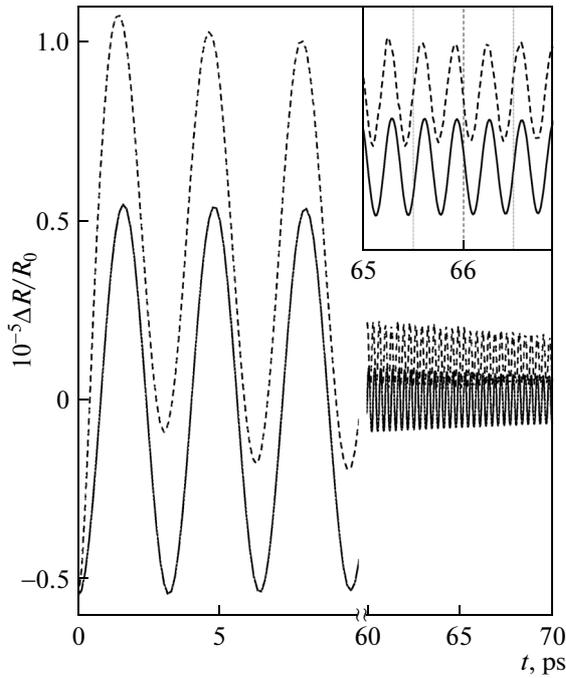

**Fig. 4.** Determination of the asymptotic phase of fully symmetric Fano phonons in the case of weak pumping. The experimental curve is shown by a dashed line, and the fitting decaying harmonic function is shown by a solid line. The inset shows, in an expanded scale, a few oscillation cycles for large time delays and the result of fitting, except the constant, that illustrate the coincidence of phases in the range of fitting.

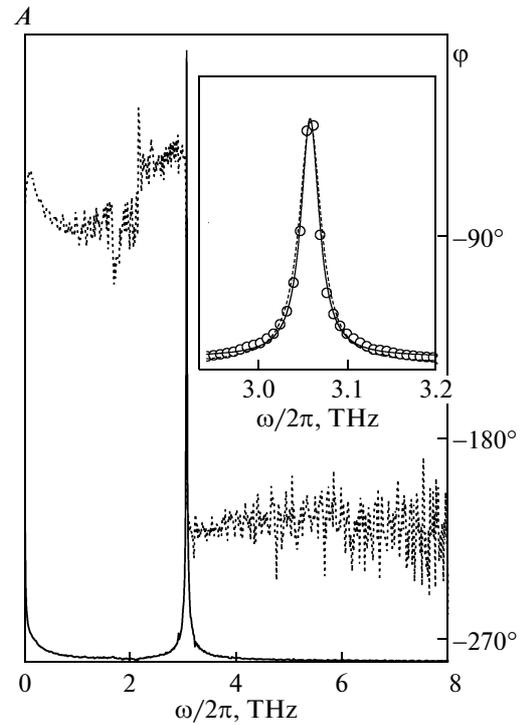

**Fig. 5.** Amplitude (solid line) and phase (dashed line) spectra of photoinduced reflection under weak excitation. The inset shows the result of fitting of a fully symmetric phonon by Fano functions (dashed line, $q = -95$, $\omega/2\pi = 3.06$ THz, and $\Gamma = 0.028$ THz) and Lorentz functions (solid line, $\omega/2\pi = 3.06$ THz, and $\gamma = 0.025$ THz); the dots represent experiment.

To determine the asymptotic initial phase φ, we approximated oscillations in the range of 50–70 ps by the sum of a decaying harmonic function and a constant,

$$A = A_0 e^{-t/\tau}\cos(\omega t + \varphi) + \mathrm{const}, \quad (3.1)$$

and then extrapolated the fitting function, except the constant, to zero time delay. To determine the zero time delay, which is necessary to measure the initial phase, we interchanged the pump and probe pulses and determined the zero point from the symmetry of two kinetics [17]. A typical result of such an inversion, which illustrates the determination of the zero delay, is shown in Fig. 7.

We should specially emphasize that the phase spectra, which exist only for coherent phonons, are more sensitive to the fluctuations of atoms compared with amplitude spectra. Indeed, while, in the latter spectra, the peak corresponding to doubly degenerate $E_g$ phonons at frequency of 2.1 THz is missing at all, in the phase spectra, there are explicit irregular phase jumps near this frequency. This is associated with the fact that, as a rule, the phase carries more information about the time shape of a signal.

Let us analyze the phase spectra in greater detail. To this end, we will vary the left boundary of the time window in which the Fourier transformation is performed. We place the left boundary of the window at the oscillation extrema for time delays greater than 5 ps, at which the fast component of the relaxation contribution almost completely vanishes. We will not vary the right boundary of the window, fixing it at 70 ps. The results of this shift of the left boundary of the time window are demonstrated in Fig. 8, in which odd numbers of phase curves correspond to the position of the left boundary of the window, which roughly coincides with the zeros, whereas the even numbers coincide with the oscillation extrema. In this case, the first extremum corresponds to the trough, whereas the second corresponds to the crest of oscillations. It is obvious that the shift of the left boundary of the time window within an oscillation period does not lead to a change in the character of the phase spectrum, which demonstrates a phase jump of $\pi$ at the resonance energy. For any position of the left boundary of the window at the frequency of doubly degenerate $E_g$ phonons, there are singularities that are missing in the amplitude spectra.

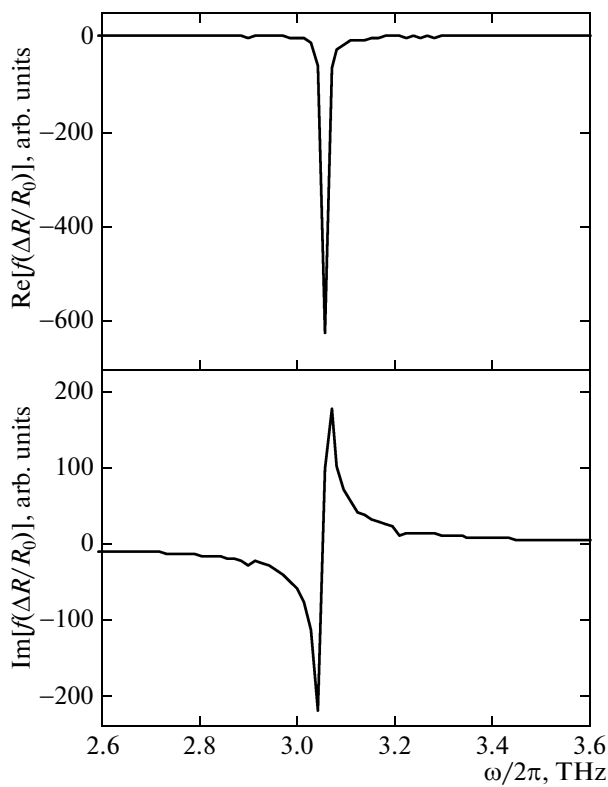

**Fig. 6.** Real and imaginary parts of the Fourier transform of the experimental signal of photoinduced reflection under weak excitation.

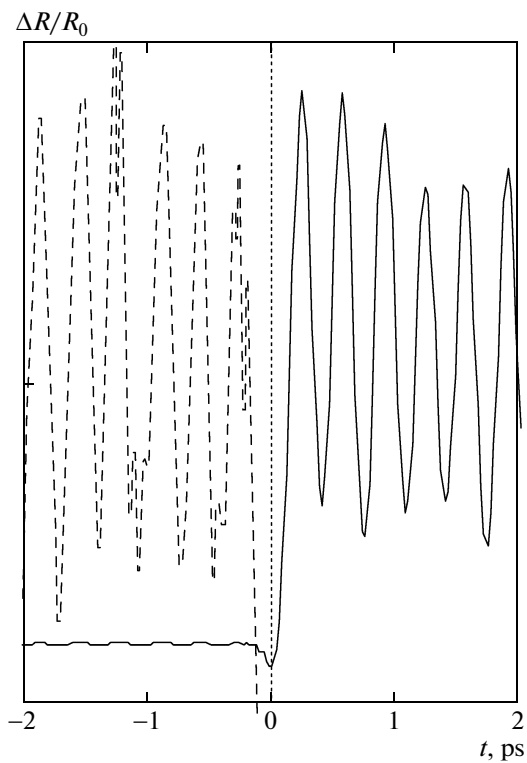

**Fig. 7.** Typical curves for the normal (solid curve) and inverted (dashed curve) succession of pump and probe pulses, that illustrate the determination of zero delay.

Thus, under weak excitation of bismuth, the oscillations of fully symmetric $A_{1g}$ phonons of bismuth have the form of a cosine function and a large value of the modulus of the asymmetry parameter, which provides evidence of kinematic excitation of these oscillations [28]. Since the asymmetry of the spectral line of these oscillations is small, one can conclude that, under weak excitation, a Breit–Wigner resonance, rather than a Fano resonance, occurs for $A_{1g}$ phonons in bismuth. Note that this result contradicts the theoretical prediction of [20], according to which a Fano antiresonance with asymmetry parameter $q \longrightarrow 0$ should occur in the case of cosinusoidal oscillations.

Increasing the excitation intensity by a factor of five, we again analyze the parameters of the photoinduced response. In this case, a decrease in the coherent amplitude is not described by a single exponential function, but is a multiexponential function of time [29, 30]. Such a decay law of the coherent state indicates that the decay probability depends on time, and the fluctuations of lattice displacements are non-Gaussian. In this case, the informativity of the power spectrum measured in the time domain does not coincide with the informativity of the spectrum of power available in the frequency domain, for example, in the case of Raman light scattering.

It follows from the experimental curve of photoinduced reflection shown in Fig. 9 that the value of the negative minimum decreases as the excitation intensity increases, whereas the relaxation and coherent amplitudes of scattering increase linearly. A decrease in the value of the negative minimum as intensity increases casts doubt on the fact that it arises due to the gradient of the electron density pressure [32], because the latter is proportional to the intensity of the exciting pulse. The time interval that separates the positions of the negative minimum and the first maximum of the signal decreases down to 240 fs compared with the case of weak excitation. The relaxation contribution varies as the excitation increases. The electron system, after reaching a state in which the variation of reflection is maximal, exhibits a decay, crossing the zero line at $t \approx 20$ ps and reaching an extremum negative value at $t \approx 50$ ps, and then monotonically approaches the equilibrium value. The coherent response is also changed: in addition to the above-mentioned modified decay law, under more intense excitation, the initial oscillations occur with a slightly increased period; however, this frequency shift is short-lived and vanishes within a few picoseconds.

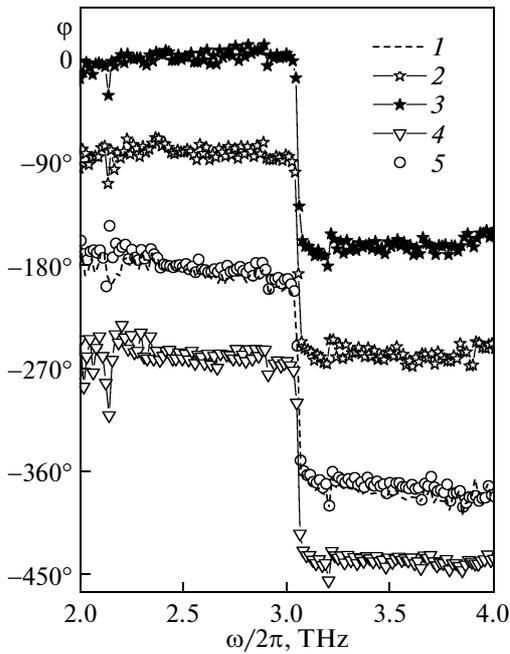

**Fig. 8.** Phase spectra under weak excitation ($F = 0.15$ mJ/cm$^2$) when the left boundary of the time window is shifted by about 5 ps. The odd numbers of phase curves correspond to the zeros, and even numbers, to the extrema of oscillations.

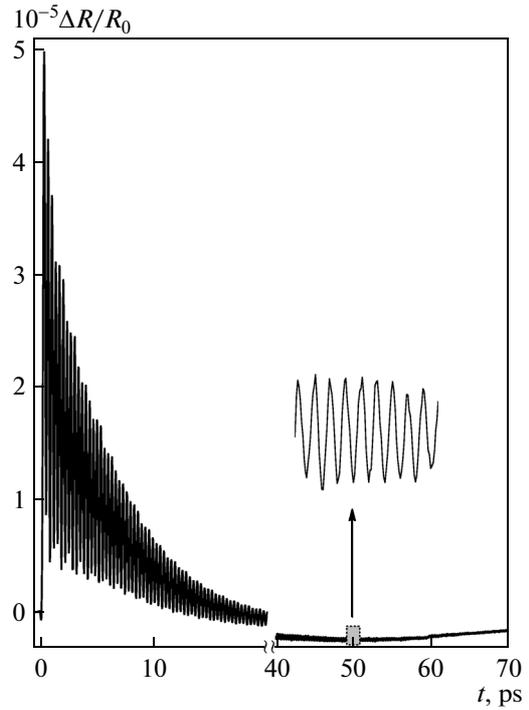

**Fig. 9.** Time-resolved normalized differential reflection $\Delta R/R_0$ of a Bi single crystal at liquid helium temperature in the case of stronger pumping ($F = 0.75$ mJ/cm$^2$). The inset shows oscillations for greater delay times in an expanded scale.

The Fourier transform of a photoinduced signal demonstrated in Fig. 10 testifies to the fact that fully symmetric $A_{1g}$ oscillations dominate in the spectrum. The shape of the spectral line of these oscillations is strongly asymmetric because the decay on the high-frequency wing occurs faster than on the low-frequency wing. The fitting of the line of fully symmetric $A_{1g}$ phonons by a Fano profile for a given excitation level yields the following parameters: $q = -3.2$, a resonance energy of 3.04 THz, and a level width of $\Gamma = 0.1$ THz. The negative value of the asymmetry parameter indicates that the minimum of the profile is reached on the high-frequency wing of the spectral line, whereas a small value of its modulus compared with that in the case of weak excitation reflects a large asymmetry of the spectral line.

To determine the asymptotic initial phase $\varphi$, we again approximate oscillations in the range of 50–70 ps by expression (3.1) and then extrapolate the harmonic function obtained to the zero point. The result of this fitting is illustrated in Fig. 11. For the large time delays at which the response is approximated, oscillations for the given excitation level have parameters (frequency and damping) that almost coincide with the parameters of oscillations for weak excitation, and their initial phase, determined by extrapolation to zero delay, is the asymptotic phase $\varphi$, close to $-4\pi/5$.

Let us compare the asymmetry parameters and asymptotic phases obtained experimentally with the theoretical curve of [20] and Eq. (2.4). Since arctangent is defined in the interval from $-\pi/2$ to $\pi/2$, we transform the asymptotic phases by adding $\pi$ to each of them, so that they belong to this interval. According to [20], such a transformation corresponds to the change of the sign of the amplitude and does not concerns the physics, because the values of phase differing by $\pi$ are equivalent. It follows from Fig. 12, which illustrates this comparison, that, for negative values of the asymmetry parameter, a decrease in the modulus of $q$, i.e., an increase in the asymmetry of a profile, leads to an increases in the asymptotic phase. This result is also observed for stronger ($F = 1.25$ mJ/cm$^2$) and weaker ($F = 0.45$ mJ/cm$^2$) excitation levels, for which the asymmetry parameters and asymptotic phases are also demonstrated in Fig. 12. For completeness of the comparison, Fig. 12 also shows the result of [20] for coherent phonons in Si.

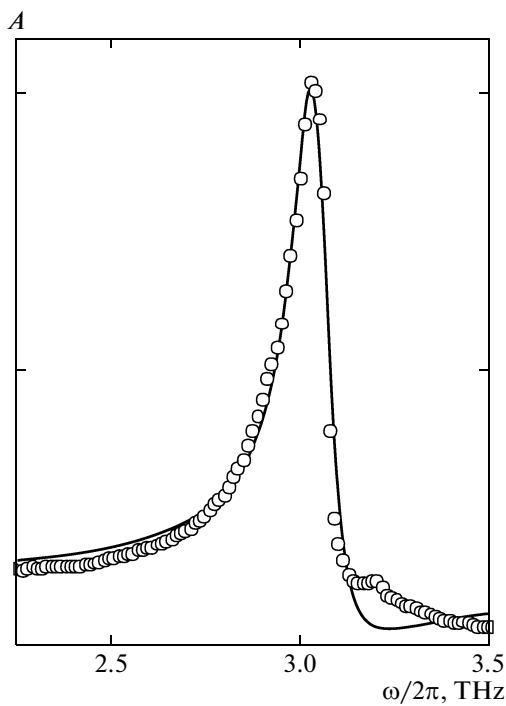

**Fig. 10.** Amplitude Fourier spectrum (dots) and its fitting by a Fano profile (solid line) in the case of stronger pumping ($F = 0.75$ mJ/cm$^2$).

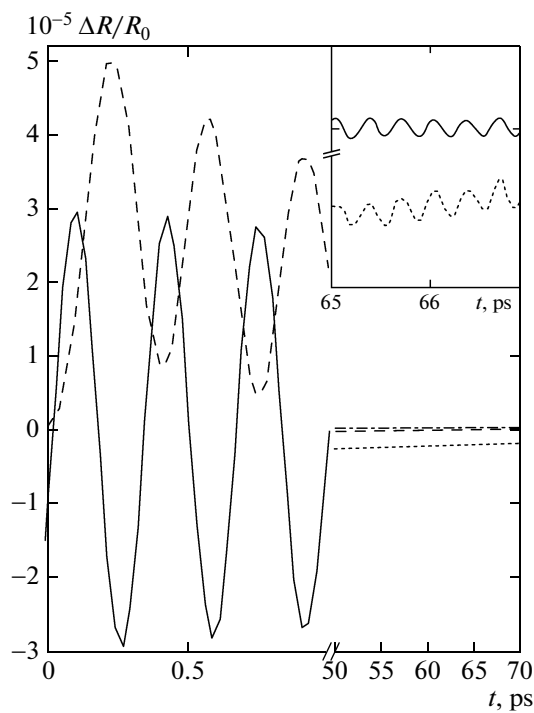

**Fig. 11.** Determination of the asymptotic phase of fully symmetric Fano phonons in the case of stronger pumping ($F = 0.75$ mJ/cm$^2$). The experimental curve is shown by a dashed line, and the fitting decaying harmonic function is shown by a solid line. The inset shows, in an expanded scale, a few oscillation cycles for large time delays and the result of fitting (except the constant) that illustrate the coincidence of phases in the range of fitting.

Thus, the experimental data provide evidence for the fact that a relationship between the asymptotic phase and the asymmetry parameter does exist; however, the functional form of this relationship does not coincide with the predictions made in [20]. The comparison of data for different excitation levels shows that, in the case of strong excitation, the asymptotic limit is reached after a few oscillations, just as in the case of coherent phonons in zinc [13], and can be attributed to the finite width of the continuum [20]. However, one cannot rule out an alternative explanation, which consists in the fact that the difference between asymptotic and experimental initial phases is associated with the frequency chirp due to the relaxation of the form of the potential [28]. Notice also that a decrease in the modulus of the asymmetry parameter with increasing excitation intensity is accompanied by a decrease in the resonance frequency and an increase in the decay rate of a coherent phonon, the latter occurring faster. This result coincides with the above-mentioned fact [12] that the real part of the eigenenergy of the coherent state of a lattice varies slower than the lifetime, which is controlled by the imaginary part of the eigenenergy.

We should specially emphasize that, for any excitation level, even for a level that leads to the destruction of the crystal, the spectrum of fully symmetric $A_{1g}$ phonons exhibits neither a Fano antiresonance, which should manifest itself as a symmetric minimum in the continuum, nor a change in the sign of the asymmetry parameter. Therefore, the verification of the predictions of the theory of [20] and our theoretical estimations by investigating coherent fully symmetric $A_{1g}$ phonons is incomplete, and, in future, one should analyze the Fano interference in the case of doubly degenerate $E_g$ phonons, for which both the above-mentioned aspects are implemented [11, 12].

To complete the investigation of coherent fully symmetric $A_{1g}$ phonons in bismuth, let us discuss the phase spectra in the case when a Fano interference is implemented. Earlier, we established that, in the case of weak excitation, when a Breit–Wigner resonance occurs, the phase spectra exhibit a steplike change of phase at the resonance frequency: the phases on the left and right of the resonance are shifted by $\pi$. Now, consider the phase spectrum in the case of the strongest excitation ($F = 1.25$ mL/cm$^2$) that was used in the present study (Fig. 13). Note at once that all the features listed below are always implemented when there exists a Fano interference. Comparing the phase spec-tra in Figs. 13 and 8, we can see that, in the case of a Fano resonance, the phase is changed almost by $2\pi$ when passing through the resonance energy, because the phases on the left and right of the resonance coincide. This is more clearly seen when the left boundary of the time window used for the Fourier transformations is shifted by 1–2 ps (see Fig. 13b). Such a shift of the left boundary of the time window to the times when the Fano resonance is already formed allows one to reveal finer details of the resonance, in particular, to localize the frequencies at which the phase jumps occur. At first glance, such behavior of the phase is unexpected. In the case when the same state can be reached by two alternative routes of which one corresponds to a narrow line for which the phase is changed by $\pi$ in a narrow spectral range from one wing of the line to the other, and the second corresponds to the background signal for which the phase in the resonance range of interest can be assumed constant, one should expect a behavior of the phase spectrum similar to the Breit–Wigner resonance.

It is obvious that the assumption that the phase of the continuum is invariant is most likely to fail because the amplitudes, rather than probabilities, are summed in the case of the Fano interference. Indeed, the comparison of the phase and amplitude spectra in Fig. 13b indicates that the first jump down of the phase occurs near the resonance frequency, whereas, a jump up of the phase roughly coincides with the position of the zero amplitude. One may assume that the singularities of the phase spectrum in the case of a Fano resonance should arise at energies determined by the roots of the denominator and the numerator of expression (1.2), i.e., at $\varepsilon = 1/q$ and $\varepsilon = -q$, when the phase is shifted in opposite directions. A comparison with the Breit–Wigner profile (1.1), for which a singularity of the phase arises only due to the pole determined by the roots of the denominator, makes obvious the difference between the phase spectra of the Fano and Breit–Wigner resonances. Here one should notice that, according to the model considered in [35], a similar behavior of the phase is shown in the case of two coupled oscillators, when the energy is pumped into the system through only one of the oscillators. One should also point out that, in the case of strong excitation, for which the amplitude and phase spectra are shown in Fig. 13b, the amplitude spectrum of doubly degenerate $E_g$ coherent phonons exhibits an explicit Fano antiresonance with $q = 0$. The behavior of this antiresonance depends on the left boundary of the time window in which the Fourier transformation is performed [11, 12]; this is obvious when comparing the spectra shown in Figs. 13a and 13b.

A profile similar to a Fano profile also arises in some models [35, 36] that differ from the Fano model. For instance, in [35], the Hamiltonian of a system consisting only of two coupled oscillators is essentially different from the models considered in [3, 20]. Nevertheless, the normalized frequency dependence of the squared modulus of the amplitude $c_1$ of oscillations of the first oscillator with damping $\gamma_1$ and circular frequency $\omega_1$ in the neighborhood of the resonance frequency $\omega_2$ of the second oscillator with negligible damping is quite similar to the Fano profile,

$$\left|\frac{c_1}{a_1}\right|^2 \approx \frac{\varepsilon^2}{\left[\varepsilon(\omega_1^2 - \omega_2^2) + \dfrac{v_{12}}{2\omega_2}\right]^2 + \gamma_1^2 \omega_2^2 \varepsilon^2}, \quad (3.2)$$

where $\varepsilon = (\omega - \omega_2)/v_{12}$, $v_{12}$ is the coupling constant. Although Eq. (3.2) differs from (1.2) algebraically and cannot be reduced to the latter, the asymmetry of the profile arises in (3.2), just as in the Fano problem, due to the mutual effect of the oscillators and is determined by the second term in square brackets in the denominator, which is proportional to the coupling constant $v_{12}$. It turns out that the external force acting on the first oscillator cannot efficiently initiate its oscillations at some frequency because this is opposed by the second oscillator. Thus, although the problem considered in [35] is, strictly speaking, not equivalent to the Fano problem, it may be useful in future when studying the interaction of various oscillatory degrees of freedom of a crystal (in the case of bismuth, the interaction of coherent phonons of $A_{1g}$- and $E_g$-symmetry). The difference of the model of [35] from the Fano model also lies in that it is impossible to change the sign of the asymmetry parameter, which is determined by the position of the resonance frequency of the oscillator through which energy is transferred with respect to the second oscillator.

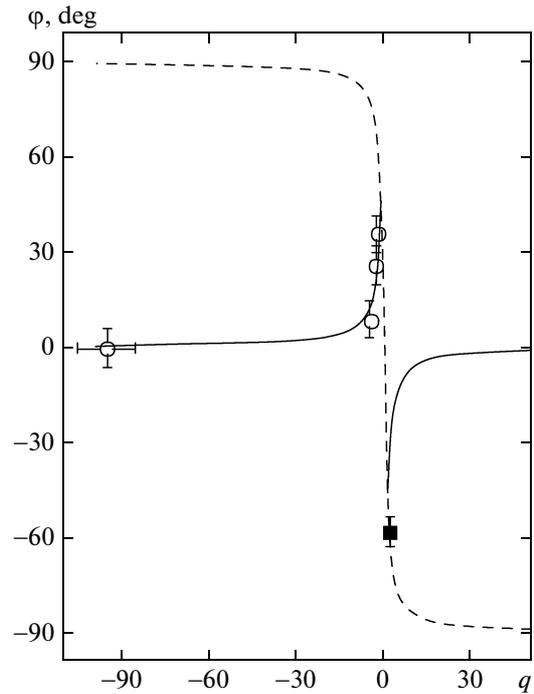

**Fig. 12.** Relationship between the asymptotic phase $\varphi$ and the asymmetry parameter $q$; (○) the values of the asymptotic phase and the asymmetry parameter of coherent $A_{1g}$ phonons for excitation densities of $F = 0.15, 0.45, 0.75,$ and $1.25$ mJ/cm$^2$, and (■) the result from [20] for coherent phonons in silicon ($q = 1.3–2.0$ and $\varphi = 58 \pm 5°$). The error in the determination of the asymmetry parameter for Si and three last excitation densities in Bi does not exceed the size of the symbol. The dashed line represents $\varphi = -\arctan q$, and the solid line, $\varphi = -\arctan(1/q)$ for the asymmetry parameter ranging from $-100$ to $45$.

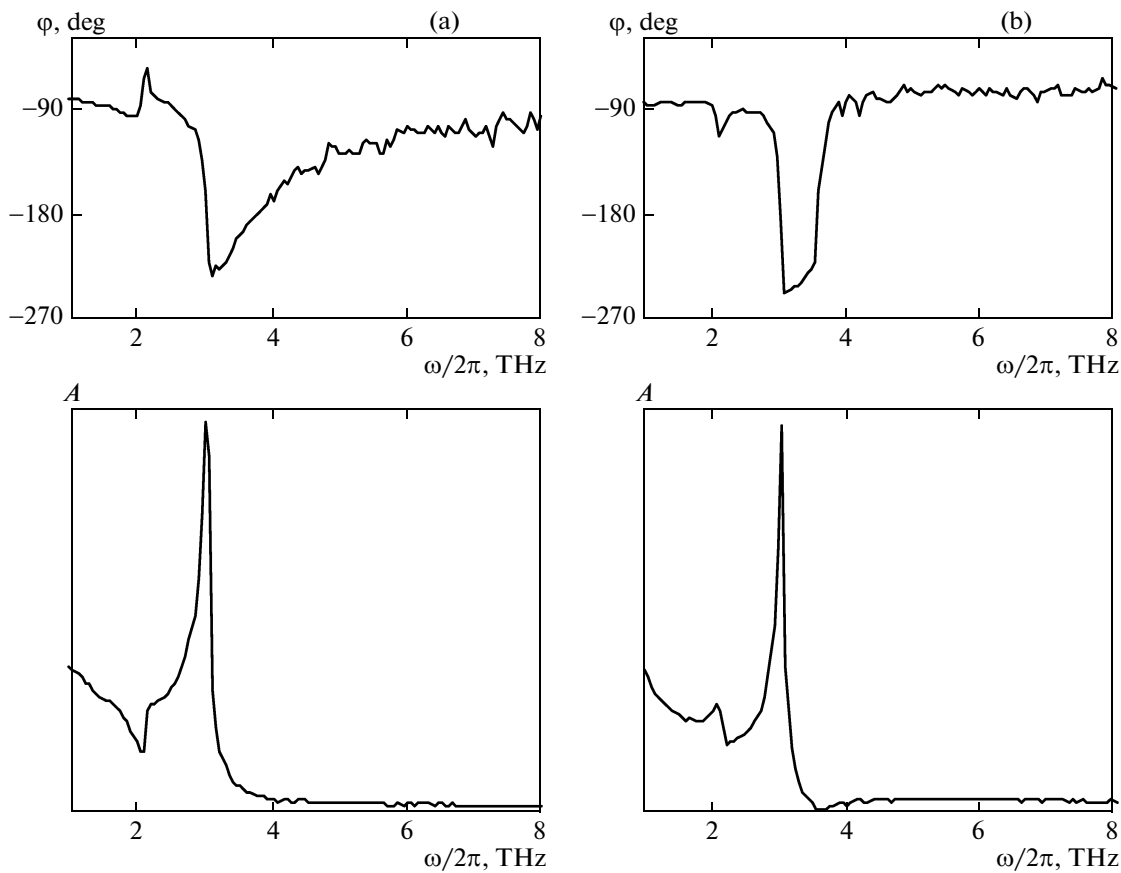

**Fig. 13.** Phase an amplitude spectra of coherent phonons in bismuth under strong excitation ($F = 1.25$ mJ/cm$^2$); (a) for a time window of 0–70 ps, and (b) for the case when the left boundary of the window is shifted by 1.78 ps and coincides with the minimum of oscillations.

## 4. CONCLUSIONS

By an example of coherent fully symmetric $A_{1g}$ phonons of bismuth excited by ultrashort laser pulses at liquid helium temperature, we have shown that, for negative values of the asymmetry parameter, the asymptotic phase increases as the modulus of the parameter decreases. Thus, we have obtained a rather strong experimental confirmation of the fact that the Fano asymmetry parameter $q$ and the asymptotic initial phase $\varphi$ of a harmonic oscillator interacting with a continuum are interrelated; however, the functional form of this relation is radically different from that predicted in [20]. According to [20], for $q \le 0$, the asymptotic phase decreases as $|q|$ decreases. Just to make sure of the correctness of the theoretical prediction about the interrelation between $q$ and $\varphi$, we should investigate the Fano interference for coherent phonons of lower symmetry, for which, under certain conditions, a Fano antiresonance occurs and the asymmetry parameter may change its sign [11, 12].

We have established that the phase spectra of coherent phonons are much more sensitive to atomic fluctuations than the amplitude spectra. For instance, in the case of weak excitation when there are no singularities in the amplitude spectra at the frequency of doubly degenerate $E_g$ phonons, the phase spectra contain irregular phase jumps at this frequency. Moreover, during the emergence of a Fano resonance, a time window arises in the phase spectrum on whose boundaries the phases coincide; this fact essentially distinguishes this resonance from the Breit–Wigner resonance, which is characterized by a single phase jump by $\pi$. The boundaries of the time window are determined by the zeros and poles of the Fano profile and roughly coincide with $\varepsilon = 1/q$ and $\varepsilon = -q$ after the formation of the resonance. Thus, in addition to the asymmetry of the spectral line that exists in the amplitude spectrum and is an integral feature of the Fano interference, we have experimentally established one more manifestation of this asymmetry, this time in the phase spectrum.

## ACKNOWLEDGMENTS

This work was supported in part by the Russian Foundation for Basic Research, project nos. 13-02-00263a and 14-02-00105.

*Translated by I. Nikitin*